\documentclass[10pt,conference]{IEEEtran}
\IEEEoverridecommandlockouts

\usepackage{type1cm}        % activate if the above 3 fonts are
\usepackage{makeidx}         % allows index generation
\usepackage{graphicx}        % standard LaTeX graphics tool
\usepackage{multicol}        % used for the two-column index
\usepackage[bottom]{footmisc}% places footnotes at page bottom
\usepackage{algorithm}
\usepackage{algpseudocode}
\usepackage{comment}
\usepackage{newtxtext}       % 
\usepackage[varvw]{newtxmath}       % selects Times Roman as basic \usepackage{tikz}
\usepackage{pgfplots}
\pgfplotsset{compat=1.18}
\usetikzlibrary{intersections, pgfplots.fillbetween}
\usepackage{subcaption}
\usepackage{cleveref}
\usepackage{enumitem,flushend}

\makeindex             % used for the subject index

\newcommand{\ba}{\mathbf{a}}

\newcommand{\e}{\mathsf{e}}

\newcommand{\jj}{\mathsf{j}}

\newcommand{\bH}{\mathbf{H}}
\newcommand{\bR}{\mathbf{R}}
\newcommand{\bx}{\mathbf{x}}
\newcommand{\bX}{\mathbf{X}}
\newcommand{\bn}{\mathbf{n}}

\newcommand{\by}{\mathbf{y}}

\newcommand{\bI}{\mathbf{I}}
\newcommand{\bu}{\mathbf{u}}
\newcommand{\bU}{\mathbf{U}}

\newcommand{\Var}{\mathbb{V}}
\newcommand{\Cov}{\mathsf{Cov}}

\newcommand{\bOmega}{\boldsymbol{\Omega}}

\newcommand{\Herm}{\mathsf{H}}
\newcommand{\Trans}{\mathsf{T}}
\newcommand{\x}{\mathsf{x}}
\newcommand{\y}{\mathsf{y}}
\newcommand{\z}{\mathsf{z}}
\newcommand{\dd}{\mathsf{d}}
\newcommand{\bigo}{\mathcal{O}}

\newcommand{\Ex}{\mathbb{E}}
\newcommand{\kk}{\kappa}

\def\bzero{\boldsymbol{0}}

\def\LOS{\mathrm{LOS}}
\def\Re{\mathrm{Re}}
\def\Im{\mathrm{Im}}

\def\tx{\mathrm{tx}}
\def\rx{\mathrm{rx}}
\def\vtx{\mathrm{vtx}}
\def\vrx{\mathrm{vrx}}
\def\NS{\mathrm{NS}}

\def\sinc{\mathrm{sinc}}
\def\max{\mathrm{max}}

\def\uc{\mathrm{uc}}

\def\NF{\mathrm{NF}}
\def\refl{\mathrm{rfl}}
\def\scat{\mathrm{scr}}

\def\Cset{\mathbb{C}}

\def\Uset{\mathbb{U}}

\def\sCN{\mathcal{CN}}
\def\Nset{\mathcal{N}}

\def\Uset{\mathcal{U}}

\definecolor{DarkGreen}{RGB}{0,150,0}
\definecolor{olivegreen}{HTML}{029A46}

\usepackage{tikz}
\usepackage{pgfplots}
\usetikzlibrary{plotmarks}
\usetikzlibrary{shapes,decorations.markings,fit,calc,patterns}
\usetikzlibrary{spy}
\usepgfplotslibrary{external}
\usetikzlibrary{external}
\newtheorem{lem}{Lemma}

\newtheorem{prop}{Proposition}
\newtheorem{corol}{Corollary}
\usepackage[noadjust]{cite} % to group references [1-3] instead of [1], [2], [3]

\def\BibTeX{{\rm B\kern-.05em{\sc i\kern-.025em b}\kern-.08em T\kern-.1667em\lower.7ex\hbox{E}\kern-.125emX}}

\newif\ifexpandabbrev
\title{\centering {Near-Field Multipath MIMO Channel Model \\for Imperfect Surface Reflection}}

\date{December 2023}
\author{
\IEEEauthorblockN{Mohamadreza Delbari$^1$, George C. Alexandropoulos$^{2,3}$, Robert Schober$^4$, H. Vincent Poor$^5$, and Vahid Jamali$^1$\\}
\IEEEauthorblockA{
$^1$Technical University of Darmstadt, Germany, 
$^2$National and Kapodistrian University of Athens, Greece,\\
$^3$University of Illinois Chicago, USA, $^4$Friedrich-Alexander-Universität Erlangen-N{\"u}rnberg, Germany,
$^5$Princeton University, USA\\
e-mails: \{mohamadreza.delbari, vahid.jamali\}@tu-darmstadt.de, alexandg@di.uoa.gr, robert.schober@fau.de, poor@princeton.edu
\vspace{-6mm}
\thanks{Delbari and Jamali’s work was supported in part by the Deutsche Forschungsgemeinschaft (DFG, German Research Foundation) within the Collaborative Research Center MAKI (SFB 1053, Project-ID 210487104) and in part by the LOEWE initiative (Hesse, Germany) within the emergenCITY center. Alexandropoulos' work was supported by the SNS JU project TERRAMETA under Grant 101097101. Robert Schober’s work was funded by the German Ministry for Education and Research (BMBF) under the program of ``Souverän. Digital. Vernetzt.'' joint project 6G-RIC (Project-ID 16KISK023) and the Deutsche Forschungsgemeinschaft (DFG, German Research Foundation) under projects SFB 1483 (Project-ID 442419336, EmpkinS). Poor's work was supported in part by the U.S National Science Foundation under Grants CNS-2128448 and ECCS-2335876.}
}
}

\begin{document}

\setlength{\hoffset}{-5mm}
\setlength{\voffset}{-5mm}
\setlength{\textwidth}{192mm}
\setlength{\textheight}{255mm}
    % Then in submission use 86%. In the name of the file, do not use space and underline!

    % Start scaling the content
%\scalebox{0.9}{
%\parbox{\textwidth}{
    
\maketitle
\pagenumbering{gobble}
\begin{abstract}
Near-field (NF) communications is receiving renewed attention in the context of passive reconfigurable intelligent surfaces (RISs) due to their potentially extremely large dimensions. Although line-of-sight (LOS) links are expected to be dominant in NF scenarios, it is not a priori obvious whether or not the impact of non-LOS components can be neglected. Furthermore, despite being weaker than the LOS link, non-LOS links may be required to achieve multiplexing gains in multi-user multiple-input multiple-output (MIMO) scenarios. In this paper, we develop a generalized statistical NF model for RIS-assisted MIMO systems that extends the widely adopted point-scattering model to account for imperfect reflections at large surfaces like walls, ceilings, and the ground. Our simulation results confirm the accuracy of the proposed model and reveal that in various practical scenarios, the impact of non-LOS components is indeed non-negligible, and thus, needs to be carefully taken into consideration.
%To compete with path loss by increasing the carrier frequency, we need extremely large reconfigurable intelligent surfaces (RISs). This leads to enlarger the NF regime so near field consideration in channel model and RIS design becomes important. This article introduces a new NF model for non-line of sight (non-LOS) links made by non-ideal reflectors. Our simulation results confirm that neglecting them in model diminishes the performance of the system. They also confirm that in some multi-user scenarios, we have to exploit reflectors to satisfy a specific performance metric.
\end{abstract}
 % End of parbox

\maketitle
\section{Introduction}
\label{Introduction}
In the field of wireless communication, reconfigurable intelligent surfaces (RISs) have recently gained considerable attention as an enabling technology for realizing programmable radio environments \cite{di2019smart,yu2021smart}. Comprising sub-wavelength elements, RISs possess the capability to dynamically change the reflected wave facilitating, e.g., the realization of virtual links between base stations (BS) and mobile users (MUs) \cite{tang2020wireless%,Basar2024
}. For the establishment of a sufficiently large link budget, RISs need to include a substantial number of elements \cite{najafi2020physics,bjornson2020rayleigh,RIS_THz_Eucap}. 
%Having a large number of cells in RIS and increasing the frequency leads to increasing the far-field distance based on $d_{\text{far-field}}=\frac{2D^2}{\lambda}$ where $D$ is the biggest dimension of the RIS and $\lambda$ is wavelength. 
Consequently, typical link distances tend to fall into the near-field (NF) regime \cite{bjornson2020power,Gong2024}, which makes it imperative to design and analyze RISs for NF communications.
 Several works have studied RIS-assisted wireless systems in the NF regime; see \cite{liu2023near} for a recent tutorial review. For example, assuming an NF line-of-sight (LOS) channel model,  the problems of beamforming and localization were studied in \cite{lu2021near} and \cite{dardari2021nlos}, respectively. Moreover, it has been shown in \cite{ramezani2023near} that in the NF regime, data multiplexing is achievable even with LOS links. However, the LOS link may not always be accessible, e.g., due to self-blockage at high frequencies. Thus, also non-LOS links have to be exploited to enhance the multiplexing gain in multi-user multiple-input multiple-output (MIMO) scenarios. Based on a point scattering model, an NF non-LOS channel model was reported in \cite{liu2023near,lu2023near}. However, experimental channel measurements suggest that dominant non-LOS components are often caused by large surfaces existing in the environment like walls, ceilings, and the ground, which all have a significant electrical aperture \cite{jansen2011diffuse,sheikh2021scattering}. We henceforth refer to them as non-reconfigurable and non-intelligent surfaces (NSs). These large NSs cannot be modeled by a point scatterer. Furthermore, reflections from NSs may lead to both reflection along the specular direction as well as non-specular directions (i.e., scattering). 
 %Interested readers are referred to \cite{liu2023near} for a recent tutorial review on NF channel models.
 
 To the best of the authors’ knowledge, a statistical NF MIMO channel model that accounts for NS has not yet been reported in the literature and is the focus of this work. Our main contributions are summarized as follows.
\begin{itemize}
    \item We develop a novel NF MIMO channel model, which captures non-LOS paths originating from imperfect reflection by large rough NSs. The model parameters are functions of the variance of roughness variations on the NS, the carrier frequency, the angles of departure (AOD), and the angles of arrival (AOA). 
    \item The non-LOS channel matrix is decomposed into the sum of a deterministic component (i.e., the channel mean) and a stochastic component. By using arguments from geometric optics, we show that regardless of the variance of the NS roughness, the deterministic component always has the form of the channel matrix of an equivalent virtual LOS link caused by an ideal surface reflection (SR). On the other hand, based on the central limit theorem (CLT), we show that the stochastic channel matrix follows a multivariate Gaussian distribution characterized by a covariance matrix. Furthermore, we drive analytical expressions for the covariance matrix for a few special cases. Here, the stochastic component captures the impact of surface scattering  (SS).
    \item We draw various design insights from the proposed model. In particular, we show that the LOS and deterministic non-LOS components of the NF model are functions of real or virtual points in the 3-dimensional space. This observation motivates the need for NF beam focusing for both LOS and NF non-LOS communication.
    \item Based on a comprehensive set of simulations, we evaluate the accuracy of the proposed channel models in an RIS case study and then verify their statistical parameters. We also demonstrate that non-LOS reflections are necessary to achieve a multiplexing gain in multi-user scenarios.
\end{itemize}

\textit{Notation:} Bold capital and small letters are used to denote matrices and vectors, respectively.  $(\cdot)^\Trans$, $(\cdot)^\Herm$, and $(\cdot)^*$ denote the transpose, Hermitian, and complex conjugate, respectively. Moreover,
%$\boldsymbol{0}_n$ and $\boldsymbol{1}_{n}$ denote a column vectors of size $n$ whose elements are all zeros and ones, respectively. 
$[\bX]_{m,n}$ and $[\bx]_{n}$ denote the element in the $m$th row and $n$th column of matrix $\bX$ and the $n$th entry of vector $\bx$, respectively. $\mathcal{CN}(\boldsymbol{\mu},\boldsymbol{\Sigma})$ and  $\mathcal{N}(\boldsymbol{\mu},\boldsymbol{\Sigma})$ denote complex and real Gaussian random vectors with mean vector $\boldsymbol{\mu}$ and covariance matrix $\boldsymbol{\Sigma}$, respectively. $\Ex\{\cdot\}$ and $\Var\{\cdot\}$ represent expectation and variance, respectively, and   $\Cov\{\cdot,\cdot\}$ denotes the covariance of two random variables. Finally, $\bigo(\cdot)$, $\mathbb{R}$, $\mathbb{C}$, and $|\Uset|$ represent the big-O notation, sets of real and complex numbers, and the Lebesgue measure of set $\Uset$, respectively. 

\section{System and Channel Models}
In this section, we present our system model and the proposed mixed-scattering NF non-LOS channel model. 

\subsection{System Model}
\label{sec: System model}
\begin{figure}[t]
    \centering
    \includegraphics[width=0.4\textwidth]{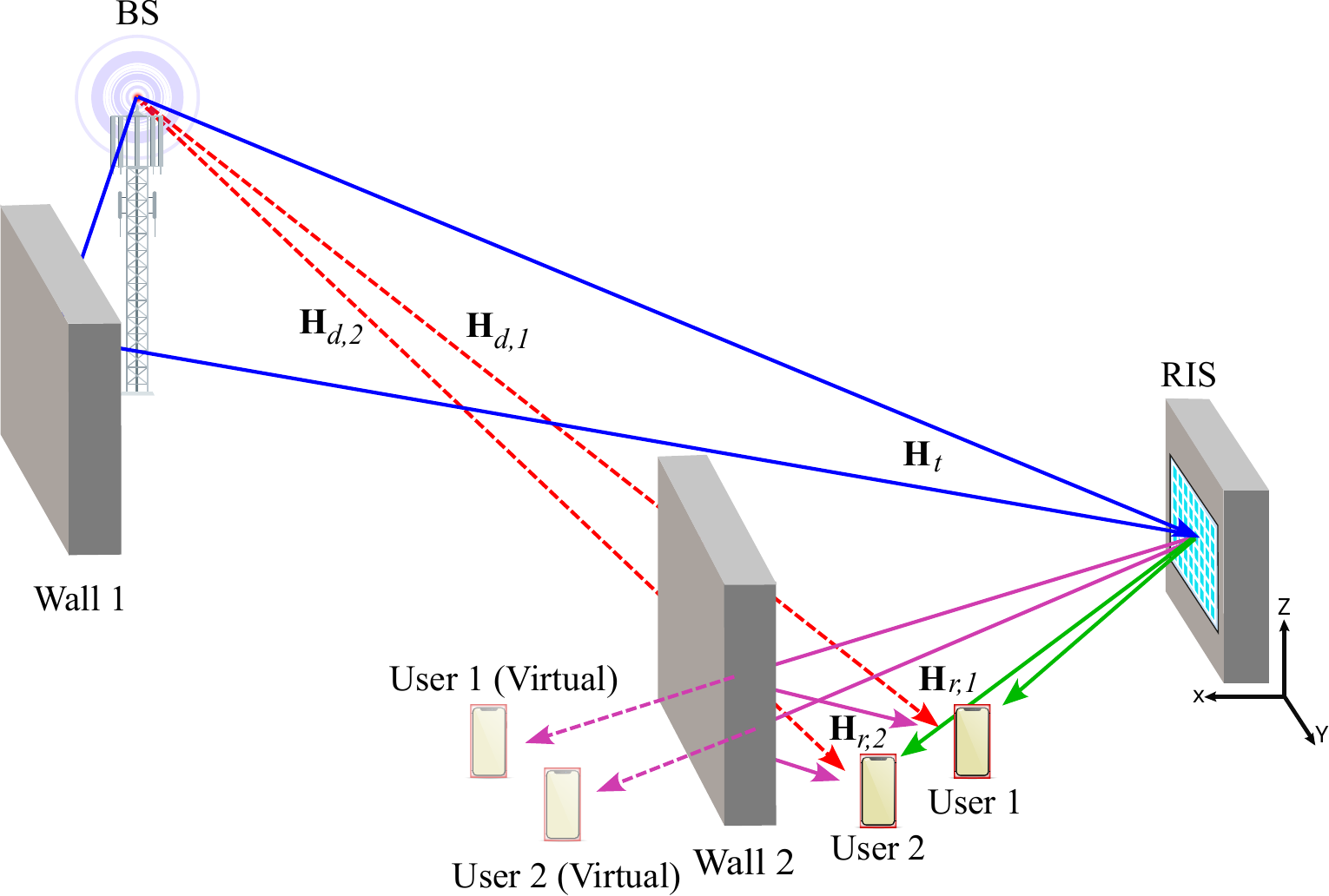}
    \caption{Schematic illustration of an RIS-assisted downlink $K$-user wireless communication system when $K=2$.}
        \vspace{-0.5 cm}
    \label{fig:system model}
\end{figure}

We consider a narrow-band downlink communication system that consists of a BS equipped with $N_t$ antennas, an RIS comprising $N$ unit-cells, and $K$ MUs each equipped with $N_r$ antennas, as depicted in Fig.~\ref{fig:system model}. The received signal vector at each $k$th user ($k=1, \dots, K$) is given by
\begin{equation}\label{Eq:IRSbasic}
	\by_k = \big(\bH_{d,k}+\bH_{r,k} \bOmega \bH_t \big)\bx +\bn_k,
\end{equation}
where $\bx\in\Cset^{N_t}$ is the transmit signal vector that 
%can be written as $\bx=\sum_{k=1}^K\bq_ks_k$, where $\bq_k\in\Cset^{N_t}$ is the beamforming vector for user $k$ and $s_k\in\Cset$ is the data symbol. Assuming $\Ex\{|s_k|^2\}=1,\,\forall k$, the beamformer 
satisfies the transmit power constraint $\Ex\{\|\bx\|^2\}\leq P_t,\,\forall k$, with $P_t$ denoting the maximum transmit power. Here, $\by_k\in\Cset^{N_r}$ and $\bn_k\in\Cset^{N_r}$ represent the received signal vector and the additive white Gaussian noise (AWGN) at the $k$th MU, respectively, i.e., $\bn_k\sim\sCN(\bzero_{N_r},\sigma_n^2\bI_{N_r}),\,\forall k$, where $\sigma_n^2$ is the noise power. Additionally, $\bOmega\in\Cset^{N\times N}$ is a diagonal matrix with main diagonal entries $\Omega_n\e^{\jj\omega_n}$, where  $\omega_n$ (with $n=1,2,\ldots,N$) represents the phase shift applied by each $n$th RIS unit-cell element and $\Omega_n$ is the unit-cell factor. The value of $\Omega_n$ depends on wavelength $\lambda$, unit-cell area $A_\uc$, the AOA, and the AOD \cite{najafi2020physics}. For simplicity, we assume a constant unit-cell factor $\Omega_n=\Omega\triangleq4\pi A_\uc\lambda^{-2},\,\forall n$  \cite{najafi2020physics} (see \cite{Wu2021} for a more detailed discussion). Furthermore, $\bH_{d,k}\in\Cset^{N_r\times N_t}, \bH_t\in\Cset^{N\times N_t}$, and $\bH_{r,k}\in\Cset^{N_r\times N}$ are the channel matrices for the BS-MU~$k$, BS-RIS, and RIS-MU~$k$ links, respectively.

For notational simplicity, in the following, we drop subscripts $d$, $t$, $r$, and $k$, and specify the channel models for a general matrix $\bH_{\NF}\in\Cset^{N_\rx\times N_\tx}$ corresponding to $N_\tx$ transmit antenna elements and $N_\rx$ receive antenna elements, and wherever necessary, explicitly refer to the  BS-MU, BS-RIS, and RIS-MU channels.

\subsection{Existing MIMO NF Channel Model}
%\vahid{For this review part, it is best if we give the existing results as briefly as possible.}

%\vahid{@Mohammadreza: Quickly give the MIMO channel model from the tutorial (+ several more references).}\\
In the NF regime, the curvature of the wavefront across the receiver (Rx) plane cannot be neglected.  Moreover, scatterers in the environment can cause multipath propagation, such that the Rx also receives signals reflected by scatterers via non-LOS paths. The following channel model has been widely used in the literature for the MIMO multipath channel in the NF regime (for example, in \cite[Eq. (25)]{liu2023near}, \cite[Eq. (8)]{lu2023near}, and \cite[Eq. (22)]{delbari2024far}):
\begin{IEEEeqnarray}{ll}\label{Eq:existingMIMO}
	\bH_{\NF} = c_0\bH_{\NF}^\LOS +
 \sum_{s=1}^S c_s\bH^\scat_s,\quad
\end{IEEEeqnarray}
where the right-hand-sided channel matrices are defined as follows:
\begin{IEEEeqnarray}{cc} 
	[\bH_{\NF}^\LOS]_{m,n} = \, \e^{\jj\kk\|\bu_{\rx,m}-\bu_{\tx,n}\|} \label{Eq:LoSnear},\\
    \bH^\scat_s=\ba_{\rx}(\bu_{s})\ba_{\tx}^\Trans(\bu_{s}),\\
 {[\ba_{\tx}(\bu_{s})]_n}   = \e^{\jj\kk\|\bu_{\tx,n}-\bu_{s}\|},
 \,\,\text{and}\,\, 
 {[\ba_{\rx}(\bu_{s})]_m}   = \e^{\jj\kk\|\bu_{\rx,m}-\bu_{s}\|}\!\!.\quad\label{Eq:nLoSnearPoint}
\end{IEEEeqnarray}

In \eqref{Eq:existingMIMO}, without loss of generality, we separate the LOS and non-LOS components of $\bH_\NF$, respectively, into scalars that determine the power of the channel and normalized matrices that determine the structure of the channel matrix. For example, $\bH_{\NF}^\LOS$ is the normalized LOS NF channel matrix and $c_0$ is the corresponding normalization factor. Here, $\bu_{\tx,n}=(x_{\tx,n},y_{\tx,n},z_{\tx,n})$ and $\bu_{\rx,m}=(x_{\rx,m},y_{\rx,m},z_{\rx,m})$ are the locations of the $n$th transmitter (Tx) antenna and the $m$th Rx antenna, respectively, and $\kk=2\pi/\lambda$ is the wave number.  Moreover,  $\ba_{\tx}(\cdot)\in \Cset^{N_{\tx}}$ and $\ba_{\rx}(\cdot)\in \Cset^{N_{\tx}}$ denote the Tx and Rx NF array response, respectively, $\bu_{s}$ is the location of the $s$th scatterer, $c_s$ denotes the normalization factor for the $s$th non-LOS path, and $S$ is the total number of scatterers. Note that the normalized LOS channel matrix can be written in terms of the Tx and Rx array responses as follows:
\begin{IEEEeqnarray}{ll} 
	\bH_{\NF}^\LOS  &= [\ba_{\rx}(\bu_{\tx,1}),\dots,\ba_{\rx}(\bu_{\tx,N_{\tx}})]\IEEEyesnumber\IEEEyessubnumber\\
 &= [\ba_{\tx}(\bu_{\rx,1}),\dots,\ba_{\tx}(\bu_{\rx,N_{\rx}})]^\Trans. \IEEEyessubnumber
\end{IEEEeqnarray}

The non-LOS channel model in \eqref{Eq:existingMIMO} is valid under the so-called point scattering assumption which presumes the scattering object is a secondary point source. Due to point scattering, the channel amplitude $c_s$ is proportional to $\frac{1}{\|\bu_{\rx}-\bu_{s}\|\|\bu_s-\bu_{\tx}\|}$ in this model, where $\bu_{\tx}=(x_\tx, y_\tx, z_\tx)$ and $\bu_{\rx}=(x_\rx, y_\rx, z_\rx)$ are the centers of the transmit and receive arrays, respectively. This observation suggests that unless the scatterer is close to either the Tx or the Rx, the impact of point-source scattering is not significant compared to the LOS link whose channel amplitude $c_0$ is proportional to $\frac{1}{\|\bu_{\rx}-\bu_{\tx}\|}$ \cite{delbari2024far}.
% For notational simplicity, we drop subscripts $d$, $t$, and $r$, and explain the channel models for a general matrix $\bH\in\Cset^{N_\rx\times N_\tx}$ corresponding to $N_\tx$ transmit antenna elements and $N_\rx$ receive antenna elements, and wherever necessary, explicitly refer to the  BS-UE, BS-RIS, and RIS-UE channels. The channel matrix $\bH$ may have both LOS and non-LOS components and hence can be modeled by Rician model as follows
% \begin{IEEEeqnarray}{ll}\label{Eq:Rician}
% 	\bH = \sqrt{\frac{K}{1+K}}\bH^\LOS + \sqrt{\frac{1}{1+K}}\bH^{\nLOS},
% \end{IEEEeqnarray}
% Here, $\bH^\LOS$ and $\bH^\nLOS$ represent the LOS and non-LOS components of $\bH$, respectively. The parameter $K$ signifies the Ricean K-factor, representing the relative power of the LOS component compared to that of the non-LOS components within the channel. In scenarios where the direct channel BS-MU is obstructed, the LOS component for this channel disappears, resulting in $K=0$. In another case, if the RIS is equipped in an environment with limited scattering, the predominant LOS component can be represented by assuming $K\to\infty$ for both the BS-RIS and RIS-MU links. Different models for $\bH^\LOS$ and $\bH^\nLOS $ in the near field regime are discussed in the following section. 

\subsection{Proposed Mixed-Scattering NF non-LOS Channel Model}
\begin{figure}[t]
    \begin{subfigure}{0.23\textwidth}
    \centering
    \includegraphics[width=1.3\linewidth]{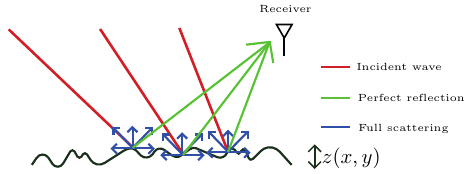}
    \caption{Reflected wave from a rough NS comprising reflection along the specular direction as well as scattering.}
        %\vspace{-0.5 cm}
    \label{fig:mixed}
    \end{subfigure}%
    \hfill
    \begin{subfigure}{0.23\textwidth}
        \centering
        \includegraphics[width=1\linewidth,height=2.5 cm]{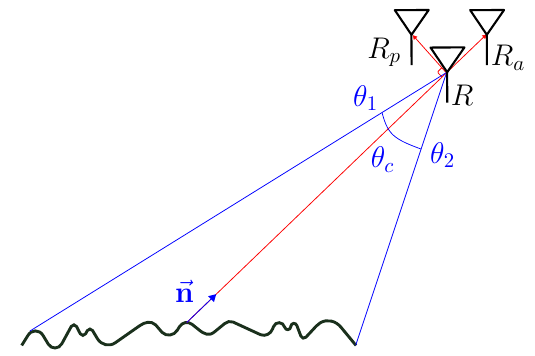}
        \caption{Vector $\vec \bn$ is aligned with $\bu_R-\bu_{R_a}$ and perpendicular to $\bu_R-\bu_{R_p}$.}
        \label{fig:mixed2}
    \end{subfigure}
    \caption{Reflected wave from an NS towards one or several receivers.}
    \vspace{-0.5 cm}
\end{figure}
In practice, the dominant non-LOS components are mainly generated by objects that have significant reflection surfaces (i.e., NS), such as walls, ground, ceiling, etc. \cite{grossman2017submillimeter}. The channel model in \eqref{Eq:existingMIMO} is not valid for such scatterers/reflectors. Thus, in this paper, we develop a statistical NF MIMO channel model that accounts for multiple NSs. The proposed channel model has the following form
%Having mentioned the non-LOS model in \eqref{Eq:existingMIMO} is not complete and we should consider also the effect of reflectors. The new total NF channel model can be written as
\begin{IEEEeqnarray}{ll}
\label{Eq:MIMO_our_model}
	\bH_{\NF} = c_0\bH_{\NF}^\LOS + \sum_{s=1}^S c_s\bH^\scat_s + \sum_{r=1}^R c_r\bH^\refl_r,\quad
\end{IEEEeqnarray}
where $\bH^\refl_r$ is the normalized non-LOS channel matrix caused by the $r$th NS in the environment and $c_r$ denotes the corresponding normalization factor. Depending on the operating frequency, the NS can be sufficiently smooth leading to ideal specular reflection, or very rough leading to non-ideal SS. In the following, we focus on modeling of each term $c_r\bH^\refl_r$ in~\eqref{Eq:MIMO_our_model}.

\section{Channel Model for Non-ideal Surface Reflection}
We are interested in characterizing the statistical properties of $c_r\bH_r^\refl$. To this end, we start by studying the physics behind the reflections from NSs. Then, we decouple $c_r\bH_r^\refl$ into two main components, namely a deterministic and a stochastic component, and characterize each one separately. Finally, we summarize the proposed model and provide insights for the design of the NF MIMO system, such as beam training and channel estimation schemes.

%There are some reasons not to have an RS for example with increasing the frequency range, the wavelength becomes shorter so it is approximately equal to the order of the surface roughness of optically smooth indoor materials. This leads to a decrease in the reflected power in the specular direction \cite{sheikh2021scattering}. We will discuss it in more detail in the next section.
\subsection{Physics Behind Non-ideal Reflection}

Various factors may impact the quality of reflection from an NS. In fact, the surface properties \cite{grossman2017submillimeter}, frequency, AOA \cite{jansen2011diffuse}, material losses, and surface roughness \cite{sheikh2021scattering} may cause scattering, diffraction, and absorption, leading to non-ideal reflection \cite{khamse2023scattering}. The key factor determining the relative importance of reflection compared to scattering is the roughness of the surface compared to the wavelength. Therefore, in this section, we start by modeling the surface roughness and analyzing the impact of variations of the roughness over NSs on the non-LOS channel matrix.

\subsubsection{Modeling of Roughness}
Assume that the NS (as shown in Fig.~\ref{fig:mixed}) lies in the $\x-\y$ plane and its height varies in the $\z$ direction. Mathematically, one may characterize the height of each point on the NS as a random variable (RV) with a given statistical distribution \cite{alissa2019wave}. We model it as a Gaussian RV and denoted it by $Z$, as is widely done in the literature \cite{jansen2011diffuse,khamse2023scattering,shi2017recovery}, i.e., $Z\sim\Nset(0,\sigma_z^2)$ with $\sigma_z^2$ representing the variance of the RV $Z$.

\subsubsection{Impact of Roughness on Reflection}
We adopt the Huygens-Fresnel (HF) principle to evaluate the impact of the NS roughness on reflection. According to this principle, each point along the beam's wavefront is the origin of a secondary spherical wave. Consequently, at any given location, the formation of the new wavefront is the result of the summation of the secondary waves \cite{goodman2005introduction,ajam2021channel}. Thus, each $c_r[\bH_r^\refl]_{m,n}$ in~\eqref{Eq:MIMO_our_model} can be modeled as follows: %$\frac{E(\bu_{\rx,m})}{E({\bu_{\tx,n}})}=$
%\begin{equation}
%\label{eq: Huygen}
%\frac{E(\bu_{\rx,m})}{E({\bu_{\tx,n}})}=\frac{\zeta}{\jj\lambda}\iint\limits_{\bu\in\Uset} \frac{\e^{\jj \kappa\|\bu-\bu_{\tx,n}\|}}{\|\bu-\bu_{\tx,n}\|}\frac{\e^{\jj \kappa\|\bu_{\rx,m}-\bu\|}}{\|\bu_{\rx,m}-\bu\|}\cos(\theta _{\tx,n})\cos(\theta _{\rx,m}){\dd}x{\dd}y,
%\end{equation}
\begin{align}
\label{eq: Huygen}
\frac{E(\bu_{\rx,m})}{E({\bu_{\tx,n}})}=\frac{\zeta}{\jj\lambda}\iint\limits_{\bu\in\Uset} &\underbrace{\frac{z_{\tx,n}}{\|\bu-\bu_{\tx,n}\|^2}\frac{z_{\rx,m}}{\|\bu_{\rx,m}-\bu\|^2}}_{\text{amplitude variations}}\nonumber\\
&\times\underbrace{\e^{\jj \kappa\|\bu-\bu_{\tx,n}\|}\e^{\jj \kappa\|\bu_{\rx,m}-\bu\|}}_{\text{phase variations}}{\dd}A,
\end{align}
where $\zeta$ is a constant ensuring surface passivity and accounting for potential losses, $\bu=(x,y,z)$ is an arbitrary point on the NS where the center of the NS is located in (0,0,0), ${\dd}A$ is the area element, and $E(\bu_{\rx,m})$ and $E(\bu_{\tx,n})$ are the electrical fields at the $m$th Rx and the $n$th Tx antennas, respectively.
Moreover, set $\Uset$ comprises the total NS area. While (\ref{eq: Huygen}) accurately models the impact of reflection, solving it for a given realization of $z(x,y)$ over the NS does not lead to a tractable model for system design and cannot provide insights. Throughout the rest of this paper, we refer to (\ref{eq: Huygen}) as the HF integral, and in Section~\ref{Simulation Result}, we numerically evaluate it for several scenarios to verify the accuracy of the proposed analytical model.

%\textcolor{blue}{\subsubsection{Amplitude approximation}
%We start with substituting $\cos(\theta_{\tx,n})$ and $\cos(\theta_{\rx,m})$ definitions in \eqref{eq: Huygen}. From Taylor series, it can be obtained $\Big|\frac{1}{\|\bu-\bu_t\|^2}-\frac{1}{u_t^2}\Big|\leq\frac{u_{\max}}{u_t^3},\,\forall t\in\{\{\tx,n\},\{\rx,m\}\}$, where $u_{\max}$ is the maximum distance of a point over NS to the center of NS, and $u_t=\|\bu_t\|$. To adopt the approximation $\frac{1}{\|\bu-\bu_{\rx,m}\|^2}\approx\frac{1}{u_{\rx,m}^2},\,\frac{1}{\|\bu-\bu_{\tx,n}\|^2}\approx\frac{1}{u_{\tx,n}^2},\,\forall \bu\in\Uset$ \cite[Eq. (4.9)]{goodman2005introduction}, $\frac{u_{\max}}{u_{\rx,n}}$ and $\frac{u_{\max}}{u_{\tx,m}}$ must be sufficiently small. Similarly, we also assume in the rest of the paper $\frac{\|\bu_{\rx}-\bu_{\rx,m}\|}{u_{\rx}}\ll1,\,\forall m$ and $\frac{\|\bu_{\tx}-\bu_{\tx,n}\|}{u_{\tx}}\ll1,\,\forall n$ hold for adopting the approximation
%\begin{equation}
%c_r[\bH_r^\refl]_{m,n}=c_r\iint\limits_{\bu\in\Uset} I(\bu,\bu_{\tx,n},\bu_{\rx,m}){\dd}x{\dd}y,
%\end{equation}
% where we defined $I(\bu,\bu_{\tx,n},\bu_{\rx,m})\triangleq\frac{1}{p|\Uset|}\e^{\jj \kappa\|\bu-\bu_{\tx,n}\|}\e^{\jj \kappa\|\bu_{\rx,m}-\bu\|}$, $c_r\triangleq\frac{\zeta_r z_\tx z_\rx}{\jj\lambda u_\tx^2 u_\rx^2}p|\Uset|$, $u_\tx=\|\bu_\tx\|$, $u_\rx=\|\bu_\rx\|$, and $p\in \Cset$ can be any constant number (we will define it properly later).}

\subsection{Proposed MIMO NF Channel Model}
We first decompose the channel matrix $c_r\bH^\refl_r$ into the following two components: 
\begin{IEEEeqnarray}{ll}\label{Eq:newMIMO}
	c_r\bH^\refl_r = c_{d,r}\bH^{d}_{r}+c_{n,r}\bH^{n}_{r},
\end{IEEEeqnarray}
where the first term is the channel mean and thus deterministic:
\begin{equation}
\label{eq: expectation_H}
    c_{d,r}\bH^{d}_{r}\triangleq\Ex\{c_r\bH^\refl_r\},
\end{equation}
while $c_{n,r}\bH^{n}_{r}$ captures the channel variations and is hence stochastic.
%First, we decompose the impact of reflection into a deterministic and a stochastic component.
%In the first step, we introduce our proposed model and discuss the impact of surface variation on the mean of the channel response.
%we derive the relative power of the coherent to the non-coherent component.
%Then, we discuss the structure of each component where we show the coherent part is deterministic while the non-coherent component follows a stochastic model. Based on our proposed model, 
%Thereby, the term $c_r\bH^\refl_r$ in (\ref{Eq:MIMO_our_model}) is expanded as
%\begin{IEEEeqnarray}{ll}\label{Eq:newMIMO}
%	c_r\bH^\refl_r = c_{d,r}\bH^{d}_{r}+c_{n,r}\bH^{n}_{r},
%\end{IEEEeqnarray}
%\textcolor{blue}{where $\bH^{d}_{r}$ and $\bH^{n}_{r}$ capture the impact of the deterministic and stochastic components}, respectively. $c_{d,r}$ and $c_{n,r}$ are the corresponding channel amplitudes, respectively.  In this model, the deterministic component is defined as the mean of the channel:
%\begin{equation}
%\label{eq: expectation_H}
%    c_{d,r}\bH^{d}_{r}\triangleq\Ex\{c_r\bH^\refl_r\}.
%\end{equation}
%Whereas, the variations in $c_r\bH^\refl_{r}$ caused by the \textcolor{blue}{stochastic} are captured by $c_{n,r}\bH^{n}_{r}$. 
In the following, we characterize the deterministic and stochastic components of $c_r\bH^\refl_r$ separately.

%As can be seen in Fig. \ref{fig:3regimes}, in case of $\sigma_z=0$, Regime 1 shows the value of normalized RS channel response; conversely, Regime 3 characterizes the SS channel response when $\kk\sigma_z$ is large enough. Finally, we propose a model from a combination of these two Regimes to follow the behaviour of Regime 2 which is the transition Regime.

%we defined $c_{d,r}\bH^{d}_{r}$ the mean of the channel ($\Ex_z\{c_r\bH^\refl_r\}$) over RV $Z$ and this concludes $\Ex_z\{\bH^{n}_{r}\}\triangleq0$.

\subsubsection{Deterministic Component}
We henceforth employ different approximations for the area element, amplitude, and phase terms in \eqref{eq: Huygen}, which are commonly used in the literature to solve complicated HF integral equations \cite{goodman2005introduction}.
%We commonly apply some approximations to the HF integral as a common way of solving it \cite{goodman2005introduction}. Amplitude variations do not affect the result of the HF integral significantly, however, phase variation is more important, therefore we have different approximations for these quantities.\\

\textbf{Area Element Approximation:} First, let us consider an ideally smooth surface, i.e., $\sigma_z=0$, for which the area element in  \eqref{eq: Huygen} becomes $\dd{A}=\dd{x}\dd{y}$. As the NS roughness ($\sigma_z$) increases, while each point $(x,y)$ on the NS still acts as a secondary source, this introduces a certain phase shift caused by $z(x,y)$. In this case, we can use approximation $\dd{A}\approx\dd{x}\dd{y}$ in \eqref{eq: Huygen} and capture the impact of the NS roughness by the phase terms in \eqref{eq: Huygen}, i.e., $\e^{\jj \kappa\|\bu-\bu_{\tx,n}\|}\e^{\jj \kappa\|\bu_{\rx,m}-\bu\|}$, which are functions of $\bu$ or equivalently $z(x,y)$ \cite{shi2017recovery}. This approximation remains valid when $z(x,y)$ is on the order of wavelengths, which is herein\footnote{We note that this approximation is not valid for arbitrary surface geometries when $z(x,y)\gg \lambda$.} the regime of interest.

\textbf{Amplitude Approximation:} 
%Let us consider \eqref{eq: Huygen}. \textcolor{blue}{Firstly, we claim that through the paper $\dd{A}\approx\dd{x}\dd{y}$ holds because $(\frac{\partial z}{\partial x})^2\ll1$ and $(\frac{\partial z}{\partial y})^2\ll1$. We start by analyzing a smooth NS then a roughness is added. The variance of this roughness is not too so our claim will hold. Secondly,} 
Using Taylor series expansion, we have $\frac{1}{\|\bu-\bu_{\tx,n}\|^2}=\frac{1}{u_\tx^2}+o\left(\frac{u_\tx^\max+u_\NS^\max}{u_\tx}\right),\,\forall \bu\in\Uset, \forall n$, where $u_\tx=\|\bu_\tx\|$. $u_\tx^\max$ and $u_\NS^\max$ are the maximum dimensions at the Tx and the NS, respectively. So $\frac{1}{\|\bu-\bu_{\tx,n}\|^2}\approx\frac{1}{u_\tx^2}$ holds when $\frac{u_\tx^\max+u_\NS^\max}{u_\tx}\ll1$. Similarly, we have $\frac{1}{\|\bu-\bu_{\rx,m}\|^2}\approx\frac{1}{u_\rx^2}$, where $u_\rx=\|\bu_\rx\|$. Hence, 
\begin{equation}\label{eq: Huygen amplitude}
c_r[\bH_r^\refl]_{m,n}=c_r\iint\limits_{\bu\in\Uset} I(\bu,\bu_{\tx,n},\bu_{\rx,m}){\dd}x{\dd}y,
\end{equation}
 where $I(\bu,\bu_{\tx,n},\bu_{\rx,m})\triangleq\e^{\jj \kappa\|\bu-\bu_{\tx,n}\|}\e^{\jj \kappa\|\bu_{\rx,m}-\bu\|}$ and $c_r\triangleq\frac{\zeta_r z_\tx z_\rx}{\jj\lambda u_\tx^2 u_\rx^2}$.\\

 \textbf{Phase Approximation:} The impact of the roughness of the NS affects the phase $\varphi(\bu)\triangleq\kappa(\| \bu_{\tx,n}-\bu\|+\| \bu_{\rx,m}-\bu\|)$ according to (\ref{eq: Huygen amplitude}).
%showing the $[\bH_r^\refl]_{m,n}$'s element related to the $n$th Tx and $m$th Rx antenna located in $ \bu_{\tx,n}$ and $ \bu_{\rx,m}$, respectively. 
This term can be simplified as follows, $\varphi(\bu)$
\begin{align}
    =&\kappa \Big(u_{\tx,n}\sqrt{1-2\frac{ \bu_{\tx,n}\cdot\bu}{u_{\tx,n}^2}+\frac{\|\bu\|^2}{u_{\tx,n}^2}}+u_{\rx,m}\sqrt{1-2\frac{ \bu_{\rx,m}\cdot\bu}{u_{\rx,m}^2}+\frac{\|\bu\|^2}{u_{\rx,m}^2}}\Big)\nonumber\\
    =&\kappa \Big(u_{\tx,n}\sqrt{1-2\frac{x_{\tx,n} x+y_{\tx,n} y+z_{\tx,n} z}{u_{\tx,n}^2}+\frac{\|\bu\|^2}{u_{\tx,n}^2}}\nonumber\\
    +&u_{\rx,m}\sqrt{1-2\frac{x_{\rx,m} x+y_{\rx,m} y+z_{\rx,m} z}{u_{\rx,m}^2}+\frac{\|\bu\|^2}{u_{\rx,m}^2}}\Big)\nonumber
\end{align}
\begin{equation}
\label{eq: orders}
\overset{(a)}{=}\!\!\kappa(u_{\tx,n}+u_{\rx,m}+z\cos(\theta_{\tx})+z\cos(\theta_{\rx})+\tilde{f}(x,y,u_{\tx,n},u_{\rx,m}))+\bigo(z^2),
\end{equation}
%where $u_\tx=\|\bu_\tx\|$ and $u_\rx=\|\bu_\rx\|$ are the distances between Tx and Rx to the center of the surface, respectively. 
where $\cos(\theta_{\tx})\approx\frac{z_\tx}{u_\tx}$ and $\cos(\theta_{\rx})\approx\frac{z_\rx}{u_\rx}$. Equality $(a)$ follows from a Taylor series expansion and the fact that $\frac{z}{u_{\tx,n}}$ and $\frac{z}{u_{\rx,m}}$ are typically small, resulting in $\frac{z}{u_{\tx,n}}\to0$ and $\frac{z}{u_{\rx,m}}\to0$. Here, $\tilde{f}$ comprises all components that are independent of $z$.
%Consider $\sigma_z^2\neq0$ but the variations are not too much such that the higher orders in (\ref{eq: orders}) affect the phase \cite{shi2017recovery}, hence, we keep the first order of $z$ and neglecting the higher orders. 
We now calculate the expected value of $c_r[\bH_r^\refl]_{m,n}$ in (\ref{eq: Huygen amplitude}) with respect to (w.r.t.) the RV $Z$ 
%exploiting (\ref{eq: orders})
\cite{shi2017recovery}, %where we adopt the approximation $\cos(\theta_\rx)=\frac{z_\rx}{\|\bu_\rx-\bu\|}\approx\frac{z_\rx}{u_\rx}$, $\cos(\theta_\tx)=\frac{z_\tx}{\|\bu_\tx-\bu\|}\approx\frac{z_\tx}{u_\tx}$  \cite[Eqs. (4-9)]{goodman2005introduction}. \textcolor{blue}{These amplitude approximations are valid because the error of it in the denominator is generally acceptably small \cite{goodman2005introduction}}. %We define $c_{r,0}\triangleq\frac{\zeta_r z_\tx z_\rx}{\jj\lambda u_\tx^2 u_\rx^2}$, 
and obtain the following expression using (\ref{eq: expectation_H}):

\begin{align}
c_{d,r}[\bH_r^d]_{m,n}&=\Ex\{c_r[\bH_r^\refl]_{m,n}\}\nonumber\approx c_r\\
\label{eq: field expected}
     &\times\kern-0.5em\iint\limits_{\bu\in\Uset}\Ex\{\e^{\jj \kappa z(\cos(\theta_{\tx})+\cos(\theta_{\rx}))}\}\times\underbrace{ \e^{\jj \kappa (u_{\tx,n}+u_{\rx,m}+\tilde{f})}}_{\kern-5em \text{Deterministic and unaffected by $z$}}\dd x\dd y.
\end{align}

%First, assume $\sigma_z^2=0$ so there is no variation on the surface. The received power at the antenna can be defined as
%\begin{equation}
%\label{eq: p_total}
%    P_\tot\triangleq\frac{1}{2\eta}\Big|\underbrace{\frac{\zeta}{\jj\lambda}\frac{1}{u_\txu_\rx}\iint\limits_{\Uset} E_{i}({\bu_\tx})\e^{\jj \kappa f(\kappa,x,y)} \cos(\theta_\tx)\cos(\theta_\rx){\dd}x{\dd}y}_{E_0(\kappa,\bu_\tx,\bu_\rx,\Uset)}\Big|^2,
%\end{equation}

%where $\eta$ is the free-space impedance, and $E_0$ is the electric field when $z=0$ in all surface points, which is a function of wavelength and positions. In this case, we assume that $P_\coh$ is coherent power and equals the power of the averaged receive electrical field over $z$. In this case, it is equal to $P_\tot$ and $P_\ncoh$ being non-coherent power is 0.

%Now 
%In the following, we neglect the effect of the second order of $z(x,y)$ in calculation which is rooted in the fact that $\kappa\frac{z(x,y)^2}{u_\rx}\ll\frac{\pi}{8}$. So it can be derived:

\begin{prop}
\label{prop: expectation}
    Assume an NS that is extremely large compared to the Rx/Tx aperture, which is a valid assumption for typical reflections from walls, ground, ceiling, etc. Moreover, assume that the RV $Z$ follows a Gaussian distribution. Based on these two assumptions, we obtain the expressions:
    \begin{IEEEeqnarray}{llll}
    \label{eq:coherent component}
&c_{d,r}(g)&=\frac{\zeta_r}{\jj\lambda}\frac{\e^{-\frac{g}{2}}}{\|\bu^r_{\vrx}-\bu_{\tx}\|}&=\frac{\zeta_r}{\jj\lambda}\frac{\e^{-\frac{g}{2}}}{\|\bu_{\rx}-\bu^r_{\vtx}\|}\IEEEyesnumber\IEEEyessubnumber\\
\big[&\bH^d_r\big]_{m,n} &= \e^{\jj\kk\|\bu^r_{\vrx,m}-\bu_{\tx,n}\|} &= \e^{\jj\kk\|\bu_{\rx,m}-\bu^r_{\vtx,n}\|},\IEEEyessubnumber
\end{IEEEeqnarray}
    %(\ref{eq: field expected}) is simplified to 
where $\bu^r_{\vtx,n}$ and $\bu^r_{\vrx,m}$ denote the virtual images of the Tx and Rx mirrored on the $r$th NS, $\bu^r_{\vtx}$ and $\bu^r_{\vrx}$ are the centers of the mirror images of the Tx and Rx arrays, respectively (see Fig. \ref{fig:system model}), and
\begin{equation}
g=\Big(\kk\sigma_z\big(\cos(\theta_{\tx})+\cos(\theta_{\rx})\big)\Big)^2.
\end{equation}
\end{prop}
\begin{IEEEproof}
\label{app: expectation}
First, we calculate the expectation of $\e^{\jj za}$ over the Gaussian RV $Z$ with $a$ being a constant. $\Ex\{\e^{\jj za}\}$ equals to:
\begin{align}
    & \int_{-\infty}^\infty \e^{\jj za} \frac{1}{\sqrt{2\pi\sigma_z^2}}\e^{-\frac{z^2}{2\sigma_z^2}} \dd z=\e^{-\frac{\sigma_z^2a^2}{2}}\nonumber\\
    &\times \underbrace{\int_{-\infty}^\infty \frac{1}{\sqrt{2\pi\sigma_z^2}}\e^{-(\frac{z}{\sqrt{2}\sigma_z}-\jj\frac{\sqrt{2}\sigma_za}{2})^2} \dd z}_{=1} \nonumber=\e^{-\frac{\sigma_z^2a^2}{2}}.
\end{align}
By replacing $a$ with $\kappa(\cos(\theta_{\tx})+\cos(\theta_{\rx}))$, (\ref{eq: field expected}) simplifies to:
    \begin{equation}
        c_{d,r}[\bH_r^d]_{m,n}=\e^{-\frac{g}{2}}c_r\iint\limits_{\bu\in\Uset}\e^{\jj \kappa (u_{\tx,n}+u_{\rx,m}+\tilde{f})}{\dd}x{\dd}y,
    \end{equation}
    %where
    %\begin{equation}
    %[c_{d,r,0}]_{m,n}=.
    %\end{equation}
 where the integral term is not a function of $z$ and assumes the same value as for perfect reflection, i.e., $z(x,y)=0,\,\,\forall x,y\in\Uset$. Hence, $[\bH_r^d]_{m,n}$ can be obtained as in \eqref{eq:coherent component} using the image theory and geometric optics \cite{a2005antenna}. 
\end{IEEEproof}

Proposition \ref{prop: expectation} shows that the mean of $c_r\bH_r^\refl$ always assumes the form of specular reflection, however, its amplitude $c_{d,r}(g)$ decreases as the NS becomes rougher, i.e., $\sigma_z$ increases. The result in Proposition \ref{prop: expectation} will be verified for practically large NSs in Section~\ref{Simulation Result} against the HF integral in \eqref{eq: Huygen}. 

\subsubsection{Stochastic Component}
Unlike the point scattering channel model provided in (\ref{Eq:existingMIMO}), the channel matrix corresponding to a rough NS is not fully deterministic.  We derive a statistical model for the NF channel in the following. 
%when $\kk\sigma_z\gg1$ (
%in Regimes 2 and 3. This finally leads to $\e^{-\frac{g}{2}}\to0$ so $c_r\bH_r^\refl\to c_{n,r}\bH_r^n$.
%We will evaluate this proposal later in Sec~\ref{Simulation Result}. 

\textbf{Joint PDF of the Elements in $\bH_r^n$ in \eqref{Eq:newMIMO}:} %Let us define $\varphi(x,y)=\kappa z(\cos(\theta_\tx)+\cos(\theta_\rx))$ in (\ref{eq: field expected})
$\varphi(\bu)$ in (\ref{eq: orders}) has a certain distribution which is identical for all $\bu\in\Uset$. 
%When $\kk\sigma_z\gg1$, we conclude $\varphi(x,y)\sim\Uset(0,2\pi)$ where $\Uset(a,b)$ represents uniform distribution between $a$ and $b$. %Because phases are fully random in this case, 
Hence, the probability density function (PDF) of the channel response in (\ref{eq: Huygen}) can be interpreted as the summation of a large number of RVs with identical distribution, which by CLT \cite{papoulis2002probability} leads to a Gaussian distribution for the elements of $\bH_r^n$. This result will be verified %in Fig. \ref{fig:distribution} 
in Section~\ref{Simulation Result}. For a Gaussian distribution, the first and second-order moments are sufficient for a full statistical characterization. Since the mean of $\bH_r^n$ is zero by our definition, we analyze the covariance matrix in the following. % for different channel coefficients.

\textbf{Covariance Matrix:} The element of the covariance matrix corresponding to the channel coefficients $c_r[\bH_r^\refl]_{n,m}$ and $c_r[\bH_r^\refl]_{n',m'}$ is denoted by $\Cov\{c_r[\bH_r^\refl]_{n,m},c_r^*[\bH_r^\refl]_{n',m'}^*\}$. Based on Proposition~\ref{prop: expectation} and using the notation $\alpha_{n,m}\triangleq c_{d,r}(g)[\bH_r^d]_{n,m}$ for brevity, yields:
%we obtain $\Cov\{c_r[\bH_r^\refl]_{n,m},c_r^*[\bH_r^\refl]_{n',m'}^*\}=\Ex\{(c_r[\bH_r^\refl]_{n,m}-c_{d,r}(g)[\bH_r^d]_{n,m})(c_r[\bH_r^\refl]_{n',m'}-c_{d,r}(g)[\bH_r^d]_{n',m'})^*\}\overset{(a)}{=}$
%\begin{align}
%\label{eq:correlation1}
%    &\Ex\Big\{|c_r|^2\iint\limits_{\bu\in\Uset} I(\bu,\bu_{\tx,n},\bu_{\rx,m})\dd x\dd y\iint\limits_{\bu'\in\Uset} I^*(\bu',\bu_{\tx,n'},\bu_{\rx,m'})\dd x'\dd y'\Big\}\nonumber\\
%    &-|c_{d,r}(g)|^2,
%\end{align}
\begin{align}
\label{eq:correlation1}
    &\Cov\{c_r[\bH_r^\refl]_{n,m},c_r^*[\bH_r^\refl]_{n',m'}^*\}\nonumber\\&=\Ex\left\{(c_r[\bH_r^\refl]_{n,m}-\alpha_{n,m})(c_r[\bH_r^\refl]_{n',m'}-\alpha_{n',m'})^*\right\}\nonumber\\
    &\overset{(a)}{=}\!\Ex\Big\{|c_r|^2\!\!\iint\limits_{\bu\in\Uset}\! I(\bu,\bu_{\tx,n},\bu_{\rx,m})\dd x\dd y\!\iint\limits_{\bu'\in\Uset}\! I^*(\bu'\!,\bu_{\tx,n'}\!,\bu_{\rx,m'}\!)\dd x'\dd y'\Big\}\nonumber\\
    &\hspace{0.43cm}-|c_{d,r}(g)|^2,
\end{align}
%. We define $p$ such that $\Ex\{[\bH_r^\refl]_{t}\}=\e^{-\frac{g}{2}}[\bH_d^\refl]_{t},$ $t\in\{(m,n),(m',n')\}$ so the spatial correlation is obtained as $[\bR]_{(n,m,n',m')}=\frac{\Cov\{[\bH_r^\refl]_{n,m}[\bH_r^\refl]_{n',m'}^*\}}{\sigma^\refl_{n,m}\sigma^\refl_{n',m'}},$ where $\sigma^\refl_{t}=\Ex\{\big|[\bH_r^\refl]_{t}-\e^{-\frac{g}{2}}[\bH_d^\refl]_{t}\big|\},\,t\in\{(m,n),(m',n')\}$ and $\Cov\{[\bH_r^\refl]_{n,m}[\bH_r^\refl]_{n',m'}^*\}=\Ex\{([\bH_r^\refl]_{n,m}-\e^{-\frac{g}{2}}[\bH_r^d]_{n,m})([\bH_r^\refl]_{n',m'}-\e^{-\frac{g}{2}}[\bH_r^d]_{n',m'})^*\}=\Ex\{[\bH_r^\refl]_{n,m}[\bH_r^\refl]_{n',m'}^*\}-\e^{-g}|[\bH_r^d]_{n',m'}|^2\overset{(a)}{=}$
where $(a)$ follows from (\ref{eq: Huygen amplitude}) and (\ref{eq:coherent component}b). In general, \eqref{eq:correlation1} cannot be simplified more unless further assumptions are made. We first define three regimes for the surface roughness, $\sigma_z$, as follows:\\
%and $I(\bu,\bu_{\tx,n},\bu_{\rx,m})\triangleq\e^{\jj \kappa(\|\bu-\bu_{\tx,n}\|+\|\bu-\bu_{\rx,m}\|)}$.\\ 
 %(see Fig. \ref{fig:3regimes}):\\
\textbf{Regime 1: $\kappa\sigma_z\ll1$}: In this case, the NS is sufficiently smooth leading to perfect SR.\\
\textbf{Regime 2:} This is a transient regime between Regimes 1 and 3, where both surface reflection and scattering are present.\\
\textbf{Regime 3: $\kappa\sigma_z\gg1$}: In this case, the NS scatters the wave in all directions leading to full SS.\\

In Regime 1, the channel is deterministic and readily obtained from Proposition \ref{prop: expectation}. For Regime 2, the HF integral in \eqref{eq: Huygen} does not lend itself to an analytical solution and must be solved numerically. However, in Regime 3, we can obtain $|c_{d,r}(g\to+\infty)|^2\to0$. Without loss of generality, we characterize the normalized covariance denoted as $[\bR]_{(n,m,n',m')}\triangleq\frac{1}{|c_r|^2|\Uset|}\Cov\{c_r[\bH_r^\refl]_{n,m},c_r^*[\bH_r^\refl]_{n',m'}^*\}$ and refer to it as spatial correlation~\cite{Arbitrary_corr}.
%define $p=1$ so $\sigma^\refl_{t}=1,\,\forall t\in\{(m,n),(m',n')\}$.
We further assume that:
%continue with the help of some other assumptions for simplifying (\ref{eq:correlation1}):

\begin{enumerate}[label=A\arabic*.]
    \item $\frac{2\|\bu_{\tx,n}-\bu_{\tx,n'}\|^2}{\lambda}<u_\tx$ and $\frac{2\|\bu_{\rx,m}-\bu_{\rx,m'}\|^2}{\lambda}<u_\rx$ hold.
    \item $\kappa\|\bu_{\tx,n}-\bu_{\tx,n'}\|\sigma_z\ll u_\tx$ and $\kappa\|\bu_{\rx,m}-\bu_{\rx,m'}\|\sigma_z\ll u_\rx$ hold.
\end{enumerate}
Note that A1 and A2 are valid in practice since the distance between nearby antennas with significant correlation is typically much smaller than the Tx-Rx distances. For ease of presentation, let us consider a local spherical coordinate system at the Rx, whose $\z$-axis passes through receive antenna locations $\bu_{\rx,m}$ and $\bu_{\rx,m'}$, and whose origin is located at $\frac{\bu_{\rx,m}+\bu_{\rx,m'}}{2}$. A similar local coordinate system is defined at the Tx w.r.t. transmit antenna locations $\bu_{\rx,m}$ and $\bu_{\rx,m'}$. Based on these assumptions and notations, we obtain the following lemma.
\begin{lem}
    \label{lem: tworxtx}
     By assuming Regime 3 under assumptions A1 and A2, (\ref{eq:correlation1}) can be simplified as follows: %$[\bR]_{(n,m,n',m')}=$
    \begin{align}
        \label{eq: lemma 1}
        &[\bR]_{(n,m,n',m')}\\&=\frac{1}{|\Uset|}\iint\limits_{\bu\in\Uset} \e^{\jj \kappa(\|\bu_{\rx,m}-\bu_{\rx,m'}\|\sin(\theta_\rx^l(\bu))+\|\bu_{\tx,n}-\bu_{\tx,n'}\|\sin(\theta_\tx^l(\bu)))}\dd x\dd y,\nonumber
    \end{align}
    where $\theta_\rx^l(\bu)$ and $\theta_\tx^l(\bu)$ are the elevation angles %of $\forall\bu\in\Uset$ \doubt 
    of point $\bu$ evaluated
    in the Rx and Tx local coordinate systems, respectively.
    %when $\bu_{\rx,m}$ ($\bu_{\tx,n}$) and $\bu_{\rx,m'}$ ($\bu_{\tx,n'}$) are assumed in $\z$-axis of a new local coordinate system , where $\bu_\rx^l\triangleq\frac{\bu_{\rx,m}+\bu_{\rx,m'}}{2}$ ($\bu_\tx^l\triangleq\frac{\bu_{\tx,n}+\bu_{\tx,n'}}{2}$) is its centre.
\end{lem}

\begin{IEEEproof}
    The proof is provided in the Appendix.%~\ref{app: tworxtx}.
\end{IEEEproof}

To derive some insights from Lemma~\ref{lem: tworxtx}, let us focus on one Tx antenna and calculate the spatial correlation among two Rx antennas located at $\bu_{\rx,m}$ and $\bu_{\rx,m'}$, denoted by $[\bR]_{m,m'}$ (note that due to reciprocity, the channel from one Rx antenna to two Txs antennas follows a similar correlation). We assume that the elevation angles $\theta(\bu)$ $\forall \bu\in\Uset$ of the points on the NS evaluated in the local Rx spherical coordinate system take value in the interval $[\theta_1,\theta_2]$; see Fig.~\ref{fig:mixed2} for an illustration.

%For the ease of presentation, similar to Lemma \ref{lem: tworxtx}, assume a local spherical coordinate system only on the Rx side, \textcolor{blue}{where its $\z$-axis passes throughout the points $\bu_{\rx,m}$ and $\bu_{\rx,m'}$, and %where $\bu_{\rx,m}$ and $\bu_{\rx,m'}$ are on the $\z$-axis,  its center is originated at $\frac{\bu_{\rx,m}+\bu_{\rx,m'}}{2}$, and $\theta$ and $\phi$ are the elevation and azimuth angles, respectively. $\theta(\bu)\in[\theta_1,\theta_2]$ $\phi(\bu)\in[\phi_1,\phi_2]$ are the corresponding value intervals $\forall \bu\in\Uset$ on the NS, see Fig. \ref{fig:mixed2} for an illustration.}
%Furthermore, assume a new spherical coordinate system when $\bu_{\rx,m}$ and $\bu_{\rx,m'}$ are on the $\z$-axis and their mid-point is the centre of the system where $\theta$ and $\phi$ show elevation and azimuth angles, respectively.
\begin{prop}
\label{prop: sinc}
    Assuming isotropic scattering within $\theta_1<\theta(\bu)<\theta_2$, the elements of spatial correlation matrix $\bR$ are given by
    \begin{equation}
    \label{eq: sinc 1}
    |[\bR]_{m,m'}|=\sinc\Big(\frac{2d_{m,m'}}{\lambda}\cos\left(\frac{\theta_2+\theta_1}{2}\right)\sin\left(\frac{\theta_2-\theta_1}{2}\right)\Big),
\end{equation}
where 
%$\theta_1$, $\theta_2$, $\phi_1$, and $\phi_2$ are constant and limit the range of the AOA. Moreover, 
$d_{m,m'}=\|\bu_{\rx,m}-\bu_{\rx,m'}\|$ and $\sinc(x)=\frac{\sin(\pi x)}{\pi x}$.
\end{prop}

\begin{IEEEproof}
    First, we substitute $\bu_{\tx,n}=\bu_{\tx,n'}$ in Lemma~\ref{lem: tworxtx} and omit indices $\rx$ and $l$ for simplicity. Transforming the integral in \eqref{eq: lemma 1} from Cartesian coordinates into spherical coordinates (i.e., $\dd x\dd y=\cos(\theta)\dd\theta\dd\phi$), we obtain the expression:
    \begin{equation}
    [\bR]_{m,m'}=\frac{1}{|\Uset|}\kern-1.5em\iint\limits_{(\phi,\theta)\in(\Phi,\Theta)} \e^{\jj \kappa\|\bu_{m}-\bu_{m'}\|\sin(\theta)}\cos(\theta)\dd\theta\dd\phi,
\end{equation}
where $(\Phi,\Theta)=\{(\phi,\theta):\phi_1\leq\phi\leq\phi_2,\theta_1\leq\theta\leq\theta_2\}$. By substituting $|\Uset|=(\phi_2-\phi_1)(\sin(\theta_2)-\sin(\theta_1))$ and continuing the integral calculation, the following result is obtained:
\begin{align} 
|[\bR]_{m,m'}|=&\Big|\int _{\phi_1}^{\phi_2} \int _{\theta_1}^{\theta_2} \frac{\e^{ \textsf {j}\frac {2\pi }{\lambda } \| \bu_{m'} - \bu_{m} \| \sin (\theta) }}{(\phi_2-\phi_1)(\sin(\theta_2)-\sin(\theta_1))} \cos(\theta)\dd\theta \dd\phi\Big|\nonumber \\ 
=&\Big|\e^{\jj \frac{\sin(\theta_1)+\sin(\theta_2)}{2}}\frac {\sin \left ({\frac {2\pi }{\lambda } \| \bu_{m'} - \bu_{m} \|\frac{\sin(\theta_2)-\sin(\theta_1)}{2}}\right) }{ \frac {2\pi }{\lambda } \| \bu_{m'} - \bu_{m} \|\frac{\sin(\theta_2)-\sin(\theta_1)}{2}}\Big|\\=&\sinc\Big(\frac{2d}{\lambda}\cos\left(\frac{\theta_2+\theta_1}{2}\right)\sin\left(\frac{\theta_2-\theta_1}{2}\right)\Big).\nonumber
\end{align}
%where $d=\| \bu_{m'} - \bu_{m} \|$, $\sinc(x)=\frac{\sin(\pi x)}{\pi x}$, and $b=\frac{\sin(\theta_2)+\sin(\theta_1)}{2}$.
This completes the proof.
\end{IEEEproof}

Proposition~\ref{prop: sinc} reveals that the correlation is maximum (minimum) if $\bu_m-\bu_{m'}$ is parallel (perpendicular) to $\vec\bn$ which is the vector pointing from the center of the NS towards the center of the Rx, see $R_a$ and $R_p$ in Fig. \ref{fig:mixed2}. Therefore, we simplify (\ref{eq: sinc 1}) for these two special cases in the following corollary.
\begin{corol}
From Proposition \ref{prop: sinc}, we have:
\label{corol: sinc}
$$
|[\bR]_{m,m'}|=
\begin{cases}
\sinc\Big(\frac{2\|\bu_R-\bu_{R_a}\|}{\lambda}\sin^2\left(\frac{\theta_c}{2}\right)\Big), \theta_1=-\frac{\pi}{2},\, \theta_2=-\frac{\pi}{2}+\theta_c,\\
\sinc\Big(\frac{2\|\bu_R-\bu_{R_p}\|}{\lambda}\sin\left(\frac{\theta_c}{2}\right)\Big),\,\, \theta_1=-\frac{\theta_c}{2},\, \theta_2=\frac{\theta_c}{2},
\end{cases}
$$
where we have defined $\theta_c\triangleq\theta_2-\theta_1$.
\end{corol}
\begin{IEEEproof}
    The proof is omitted here due to the space limitation but can be obtained by substituting $\theta_1$ and $\theta_2$ with their definitions and using further trigonometric manipulations.
\end{IEEEproof}

 Corollary \ref{corol: sinc} offers the following insights. Firstly, as $\sin^2(x)<\sin(x),\,\,\forall x\in(0,\pi)$, the correlation between aligned antennas ($\bu_m-\bu_{m'}$ parallel to $\vec\bn$) is larger compared to the perpendicular case ($\bu_m-\bu_{m'}$ perpendicular to $\vec\bn$) when $\|\bu_R-\bu_{R_a}\|=\|\bu_R-\bu_{R_p}\|$. Secondly, increasing the range of the AOA ($\theta_c$) reduces the correlation between antennas. This occurs when the NS is larger or the antennas are closer to the NS. The accuracy of the results in the Corollary~\ref{corol: sinc} will be verified in Section~\ref{Simulation Result}.

\textbf{Channel Power Gain:} %The channel power can be derived when $\|\bu_m-\bu_{m'}\|=0$ in (\ref{eq: sinc 1}) which is $\Ex_z\{|c_{n,r}|^2\}=|\Tilde{h}_r^d|^2$ and 
Based on Proposition~\ref{prop: expectation}, we can directly calculate the channel power gain in Regime 1 since the channel is deterministic. In Regime 3, the channel power gain can be derived analytically using the law of conversation of energy and exploiting the symmetry of full scattering. In fact, in this regime, the deterministic component is zero, e.g., $\e^{-\frac{g}{2}}\to0$ in (\ref{eq:coherent component}), and we argue that the reflected channel power gain is distributed in all directions equally without any preference due to homogeneous SS. We define the channel power gain reflected from 
the $r$th NS as $c_{n,r,+\infty}$, which can be derived as follows
\begin{equation}
    |c_{n,r,+\infty}|^2\triangleq\frac{P_\rx}{P_\tx}=\frac{\zeta_r}{\lambda}\Big(\frac{A_\rx D_r}{4\pi u_\rx^2}\Big)\Big(\frac{A_r D_\tx}{4\pi u_\tx^2}\Big),
\end{equation}
where $P_\rx$ ($P_\tx$) is the received power (transmitted power), $A_\rx$ ($A_r$) is the Rx ($r$th NS) effective area, and $D_\tx$ ($D_r$) is the Tx ($r$th NS) directivity. We assume $D_r=2$ (i.e., 3~dB gain) since NS uniformly reflects the wave \textit{only} in half of the space. Unfortunately, for Regime 2, there is no closed-form solution, and instead (\ref{eq: Huygen}) needs to be solved. Since this task is not analytically tractable, we model the channel power gain of the stochastic component as follows: 
\begin{equation}
\label{eq: stochastic power}
    \Ex\{|c_{n,r}(g)|^2\}=(1-\e^{-\frac{g}{2}})^2|c_{n,r,+\infty}|^2,
\end{equation}
which is consistent with the limiting values of the channel power gain in Regimes 1 ($g=0$) and 3 ($g\to+\infty$). We will show via simulations in Fig. \ref{fig:3regimes} that this heuristic power model closely follows the results obtained by numerically solving the HF integral in \eqref{eq: Huygen}.
%\begin{equation}
%    \Ex\{|c_{n,r}|^2\}=(1-\e^{-\frac{g}{2}})|\Tilde{h}_r^d|^2.
%\end{equation}
\subsection{Summary and Discussion}
%\vahid{@Mohammadreza: Here, we give the abstract model. Discuss how such a channel can be estimated in practice. Why using it is important at least for performance analysis, etc.}

The proposed overall NF channel model is summarized as follows
%(\ref{Eq:MIMO_our_model_location}) where the reflection channel model has been written in (\ref{Eq:newMIMO}).  
\begin{align}
\label{Eq:MIMO_our_model_location}
	\bH_{\NF} = &c_0\underbrace{\bH_{\NF}^\LOS(\bU_0)}_{\text{deterministic}} + \sum_{s=1}^S c_s\underbrace{\bH^\scat_s(\bU_s)}_{\text{deterministic}}  \nonumber\\ &+\sum_{r=1}^R \Big(c_{d,r}\underbrace{\bH^d_r(\bU_r)}_{\text{deterministic}}+c_{n,r}\underbrace{\bH^n_r}_{\text{stochastic}}\Big),\quad
\end{align}
where $\bU_0=[\bu_\rx, \bu_\tx]$, $\bU_s=[\bu_\tx,\bu_\rx,\bu_s]$, and $\bU_r=[\bu_\tx,\bu_\rx,{\bu^r_\vrx},{\bu^r_\vtx}]$. Interestingly, the generalized MIMO NF channel model in (\ref{Eq:MIMO_our_model_location}) indicates that all channel components are functions of real or virtual locations of Tx and Rx. This has profound implications in practice for efficient NF beam training \cite{delbari2024far,liu2023near,lu2023near,ramezani2023near,NF_beam_tracking}, where essentially the Tx and Rx learn through NF beam training to focus on the points $\bu_\tx$, $\bu_\rx$, ${\bu^r_\vrx}$, and ${\bu^r_\vtx}$~\cite{George2022}. Once the NF beamformer leading to a sufficiently high channel power gain is found, the end-to-end channel can be estimated by conventional pilot-based estimators \cite{jamali2022}.

%In other words, the proposed model suggests that it is enough for Tx and Rx to examine a set of NF codewords during the initial beam training (i.e., focusing on  

%\textcolor{red}{Therefore, despite the simplifying assumptions made for the derivative of the proposed NF channel model, the estimated channel (after beam training) would accurately characterise the end-to-end channel based on this approach.}

%the most important parameters in estimating the channel for the Tx are $\bu_\rx$, $\bu^r_\vrx$, $\bu_\tx$ and $\bu^r_\vtx$ for all reflectors (especially one has large $\rho_\coh$ value). Both $c_r$ and $c_s$ are also related directly to these values. Hence by estimating these values, the end-to-end NF channel model can be derived. In summary, the end-to-end channel model depends only on positions. Although there are many imperfect reasons in practical scenarios that lead to a decrease in the performance of the system compared to what we expect, exploiting these locations' information aids us in increasing multiplexing gain even if it will not reach the ideal values. With a beam training in the NF regime based on all the locations, we can affect all non-ideal factors in end-to-end channel model.

\section{Performance Evaluation}
In the following, we first verify the proposed models with simulations obtained by solving the HF integral, and subsequently, consider a scenario where an RIS serves two users via both LOS and non-LOS links.
\subsection{Model Verification}
\label{Simulation Result}
\textbf{Simulation Setup:} We assume there is a $3\times3$~m$^2$ NS located in the $\x-\y$ plane where $Z\sim\Nset(0,\sigma_z^2)$.
%and $\kappa\sigma_z$ ranges from 0 to 2 in Fig. \ref{fig:3regimes} while is equal to 3 in Fig. \ref{fig:corr} . This value is explicitly mentioned for each line in Fig. \ref{fig:distribution}. 
The carrier frequency is $28$~GHz  (i.e., $\lambda\approx1.1~$cm), which implies that the far-field of the NS starts from $\frac{2D^2}{\lambda}=3288~$m, where $D$ is the largest dimension of the NS. Moreover, the Tx is located at $(0,0,90)$, and we consider one Rx whose position is specified for each figure respectively. The NS passivity factor $\zeta=1$ is adopted for all figures.

In the following, we provide various aspects for the verification of our NF MIMO channel model.

\begin{figure}
    \centering
    \centering
\includegraphics{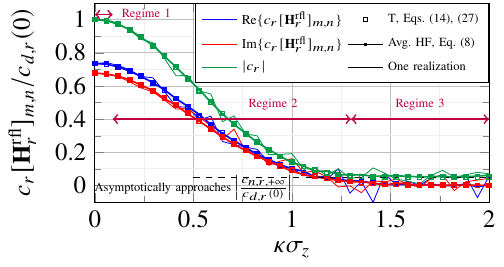}
    \caption{One realization, average, and theoretical results for the real part, the imaginary part, and the absolute value of the channel amplitude $c_r[\bH_r^\refl]_{m,n}$ normalized to ${c_{d,r}(0)}$. Based on the value of $\kappa\sigma_z$, the three different regimes, namely Regime 1 (SR), Regime 2 (transient regime), and Regime 3 (SS), are depicted. Here, T indicates the \underline{t}heoretical result.}
    \label{fig:3regimes}
    \vspace{-0 cm}
\end{figure}

\textbf{Verification of $\Ex\{c_r\bH_r^\refl\}$ and $\Ex\{|c_r|\}$ in Proposition \ref{prop: expectation} and (\ref{eq: stochastic power}):}
%From Figs. \ref{fig:distribution} and \ref{fig:corr}, it can be seen that based on the positions of receive antennas and roughness of the reflector, the correlated channel model can be introduced where the level of the roughness reaches the wavelength order, and the extension of the reflector is enough large, in the NF regime, the receive signals are approximately decorrelated when the reflected signal comes perpendicular to Rx area.
Fig. \ref{fig:3regimes} shows one realization of $\Re/\Im\{c_r[\bH_r^\refl]_{m,n}\}$, $|c_r|$ and the average (Avg.) over 100 normalized samples, as obtained by solving the HF integral in \eqref{eq: Huygen} numerically and shown with solid lines. In addition, theoretical results are shown based on Proposition \ref{prop: expectation} and (\ref{eq: total power}) depicted by square marker. Both the HF integral and theoretical results are normalized to $c_{d,r}(0)$. The curve for $\Ex\{|c_r|\}$ is obtained using (\ref{eq: stochastic power}) as follows:
\begin{equation}
\label{eq: total power}
    \frac{\Ex\{|c_r|\}}{{c_{d,r}(0)}}=\sqrt{\e^{-g}+(1-\e^{-\frac{g}{2}})^2\frac{|c_{n,r,+\infty}|^2}{|c_{d,r}(0)|^2}}.
\end{equation} 

As can be seen from Fig.~\ref{fig:3regimes}, the averaged results obtained from the HF integral (solid lines) are in perfect agreement with the proposed theoretical results (square markers). The average of both the real and imaginary parts of $\frac{c_r[\bH_r^\refl]_{m,n}}{c_{d,r}(0)}$ derived from the HF integral in \eqref{eq: Huygen} have identical slopes $\e^{-\frac{g}{2}}$ derived from $\frac{c_{d,r}(g)}{c_{d,r}(0)}$ in (\ref{eq:coherent component}a). Moreover, the average of $\frac{|c_r|}{c_{d,r}(0)}$ obtained from the HF integral asymptotically converges to $\frac{|c_{n,r,+\infty}|}{|c_{d,r}(0)|}$ when $\kappa\sigma_z$ is sufficiently large.
%, while the absolute value of the channel response asymptotically converges to the square root of the proposed channel power, denoted by $|c_{n,r,+\infty}|$.}

\begin{figure}
\centering
    \includegraphics{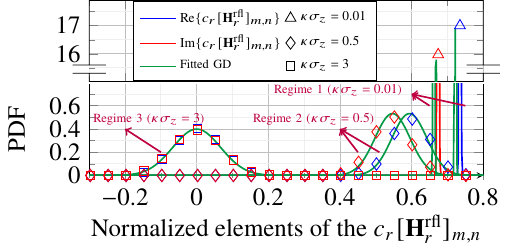}
    \caption{The distribution of the real and imaginary parts of $c_r[\bH_r^\refl]_{m,n}$ obtained with the HF integral in (\ref{eq: Huygen}) normalized by $c_{d,r}(0)$.}
    \label{fig:distribution}
    \vspace{-0.4 cm}
    \end{figure}

\textbf{Verification of the PDF of the Elements in $\bH_r^n$:} Fig. \ref{fig:distribution} shows the PDF of $\Re/\Im\{c_r[\bH_r^\refl]_{m,n}\}$ for $\kk\sigma_z=0,\,0.5,\,3$ obtained with the HF integral in \eqref{eq: Huygen}. The histograms obtained by numerically solving the HF integral in \eqref{eq: Huygen} for various realizations of $z(x,y)$ are in perfect agreement with the Gaussian distribution in all three regimes. This validates the adoption of the Gaussian distribution.

\begin{figure}
        \centering
\includegraphics{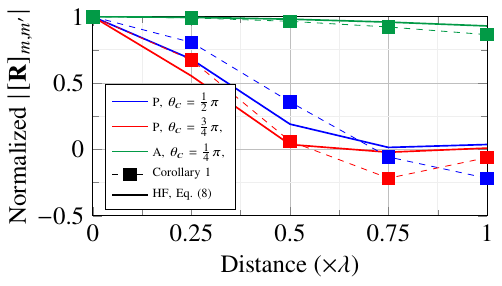}
        \vspace{-0.1 cm}
        \caption{Comparison of the normalized spatial correlation for different positions of the Rx antennas. For conciseness, we use the abbreviation (P, A), indicating %\underline{s}imulation result derived from the HF, \underline{t}heory result from Corollary~\ref{corol: sinc}, 
        \underline{p}erpendicular and \underline{a}ligned Rx antennas w.r.t. $\vec{\bn}$, respectively. For all results, $\kappa\sigma_z=3$ was used.}
        \label{fig:corr}
        \vspace{-0.2 cm}
    \end{figure}

\textbf{Verification of the Correlation of the Elements of $\bH_r^n$ in Corollary \ref{corol: sinc}:} Figure~\ref{fig:corr} depicts the spatial correlation between two Rx antennas as a function of their distance. As can be seen, the results of Corollary \ref{corol: sinc} closely follow the trends of the HF integral approximately. The spatial correlation depends on $\sigma_z$, $\theta_c$, and the antennas' positions. We study two particular cases considered in Corollary~\ref{corol: sinc} where $\vec{\bn}\perp\bu_R-\bu_{R_p}$ and $\vec{\bn}\parallel\bu_R-\bu_{R_a}$, respectively, see Fig. \ref{fig:mixed2}. In the first case, the spatial correlation decreases faster with increasing distance than in the second case verifying the results of Corollary~\ref{corol: sinc}. 
%For the first case, we changed $\theta_c$ (in \cite{bjornson2020rayleigh,jamali2023impact}, authors assumed $\theta_c=\pi$ which is a particular case in Corollary~\ref{corol: sinc})}.
Moreover, as Corollary \ref{corol: sinc} suggests, we observe that the correlation between the antennas increases for smaller $\theta_c$.
    
    % Add some vertical spacing between the figures
    %\vspace{1cm}
    
    %\caption{HF verifies the theory results of mean, PDF and spatial correlation.}
    %\vspace{-0.5 cm}
    %\label{fig:model verification}
%\end{figure}

\subsection{Impact of Scattering in a Multi-User Scenario} 
To investigate the impact of NS on NF MIMO communications, we first present the considered setup and benchmark schemes. Subsequently, we present our simulation results.\\

\textbf{Simulation Setup:} We employ the simulation setup for coverage extension illustrated in Fig. \ref{fig:system model}.
%We assume there is one user located in a random position in area $\bu_\text{MU}\in\{(\x,\y,\z): \x=30~\text{m}, -6~\text{m}\leq \y \leq -5~\text{m}, -6~\text{m}\leq \z \leq -5~\text{m}\}$ in Fig. \ref{fig:CDF} and 
We assume there are $K=2$ users located at $[23,-23,-5]$ and $[28,-28,-5]$. Moreover, we assume two walls with the following location parameters; wall 1: $\x=45$~m, $30$~m $\leq \y\leq 40$~m, $-5$~m $\leq\z\leq5$~m, wall 2: $\x=30$~m, $-27$~m $\leq \y\leq -17$~m, $-5$~m $\leq\z\leq5$~m, and we choose $\zeta$ to obtain a $3$~dB loss for each wall when it reflects in specular direction. The second wall severely blocks the BS-MU channel by $-40$~dB. The BS comprises a $4\times4$ uniform planar array (UPA) centered at $[40,40,5]$~m parallel to the $\x-\z$ plane. The RIS comprises a UPA located at $[0,0,0]$~m, consisting of $N_\y\times N_\z$ square tiles along the $\y$- and $\z$-axes, respectively, where $N_\y=N_\z=100$. The element spacing for the UPAs at both the BS and RIS is half a wavelength. The MUs have a single antenna ($N_r=1$). The noise variance is computed as $\sigma_n^2=WN_0N_{\rm f}$ with $N_0=-174$~dBm/Hz, $W=20$~MHz, and $N_{\rm f}=6$~dB. Moreover, the Ricean factor, which is the relative power of the LOS links to the point scattering ones, is set as $K_f=(-100,100,8)$~dB for the BS-MU, BS-RIS, and RIS-MU channels, respectively. The center frequency is $28$~GHz, the reference path loss at $d_0=1$~m is $\beta=-61$~dB, and the path-loss exponent is $\eta = 2$.  
%Transmit power %is assumed $P_t=5$~dBm for Fig. \ref{fig:CDF} and changes in Fig. \ref{fig:SINR} to achieve desired signal-to-noise-ratio (SNR). 
%\textcolor{blue}{We assume a linear beamforming in the BS.}

%\textbf{Impact of scattering in a single user scenario:} Fig.~\ref{fig:CDF} shows the SNR of the user in three different cases with and without considering the wall reflection effect assuming a loss-less and lossy wall. In all cases, RIS phase shifts are designed to maximize the received SNR based only on user location. When the impact of the wall is neglected, Fig. \ref{fig:CDF} suggests that the user with SNR threshold ($\SNR_\thr$) below 12~dB can always be guaranteed. However, when the impact of the wall is taken into account, the received SNR significantly vary from 0 to 20-40~dB depending on the loss parameter of the wall. Interestingly, in almost 50\% of situations, the SNR of the user is lower than the required 12~dB which is required to send and receive data. As a result, neglecting the reflection from the walls may lead to an unrealistic communication setting in practice.
%\begin{figure}[t]
%    \centering
    %\includegraphics[width=0.45\textwidth]{Figures/CDF_new.pdf}
%    \caption{SNR comparison when phase shifts are designed to focus on user location with and without assuming the effect of wall.}
%        \vspace{-0.6 cm}
%    \label{fig:CDF}
%\end{figure}

\begin{figure}[t]
    \centering
    \includegraphics[width=0.45\textwidth]{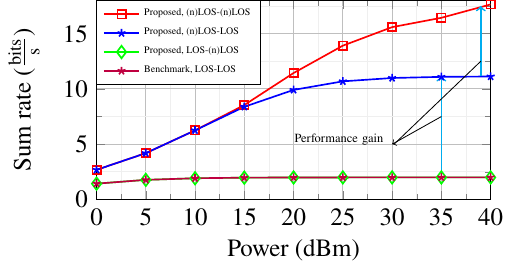}
    \caption{Sum rate vs. transmit power for $K=2$.}
        \vspace{-0.5 cm}
    \label{fig:SINR}
\end{figure}

%\mohamadreza{Alternative interesting simulation could be illuminating an area and then see the minimum SNR in that area after applying the effect of the wall reflection}
\textbf{Performance Evaluation:} Figure~\ref{fig:SINR} shows the achievable sum rate versus (vs.) the BS transmit power for two users. We adopt linear beamforming, where the BS transmits along the LOS and/or the non-LOS BS-RIS channel paths. Similarly, the RIS is divided into $10$ tiles, where a given tile reflects the wave along the LOS or the non-LOS RIS-MU channel paths. Moreover, we adopt the algorithm proposed in \cite{jamali2023impact} to optimize the BS and RIS beamformers. To show the benefits of exploiting non-LOS paths for beamforming, we distinguish four cases: \textit{i) (n)LOS-(n)LOS:} The BS and RIS transmit along the LOS and non-LOS paths in both the BS-RIS and RIS-MU channels, respectively. \textit{ii) (n)LOS-LOS:} The RIS reflects the wave only along the LOS RIS-MU paths. \textit{iii) LOS-(n)LOS:} The BS transmits only along the LOS BS-RIS path. \textit{iv) LOS-LOS:} The BS and RIS only exploit the LOS paths for beamforming, which we consider as a benchmark. 
%We adopt both LOS+nLOS links in the proposed methods at least in one of the paths BS-RIS or RIS-MU and show it with (n)LOS. Hence there are three different possibilities. First, both LOS and nLOS links are exploited in both paths, i.e., (n)LOS-(n)LOS. Second, (n)LOS are exploited only in the BS-RIS path and the RIS-MU path is only served by the LOS link, i.e., (n)LOS-LOS. Third, only the LOS link is used in the BS-RIS path and (n)LOS are exploited in the RIS-MU path, i.e., LOS-(n)LOS. As a benchmark, we consider the case where only the LOS links are exploited by BS-RIS and RIS-MU, i.e., LOS-LOS. 
From Fig.~\ref{fig:SINR}, we observe that the performance of the benchmark quickly saturates as the multi-user interference dominates the noise for both users. Moreover, we observe that using the non-LOS paths of only the RIS-MU channels does not improve the performance suggesting that the BS-RIS link is the bottleneck for achieving a multiplexing gain. This observation is confirmed by the performance improvement achieved when the non-LOS  BS-RIS path is exploited for beamforming. Here, in the NF regime,  the RIS with its large aperture is able to simultaneously focus on both MUs, despite the fact that they are located in the same direction from the RIS. A further gain can be achieved when the non-LOS paths of both the BS-RIS and the RIS-MU channels are exploited for beamforming, which is due to the additional spatial diversity.  %However, if RIS uses only the LOS link while BS exploits all (n)LOS links, the performance outperforms that of previous schemes in this case because the dimension of RIS is relatively large and is able to distinguish the users although they are in the same direction relatively. Compared with other schemes, the proposed scheme which exploits non-LOS links on two sides can generate a full-rank end-to-end channel and achieves multiplexing gain via spatial domain. 
In summary, Fig. \ref{fig:SINR} underlines the importance of exploiting the non-LOS components created by NS in multi-user scenarios.

%We adopt  LOS+non-LOS links in the proposed method to show the benefit offered by the spatial domain. As a benchmark, we consider the case where only the LOS links to the RIS are used. Furthermore, to show the relative benefit of diversity in the  by exploiting non-LOS links, we consider two variants for both the proposed and the benchmark schemes, based on whether the power domain is exploited, namely non-orthogonal multiple access (NOMA) and treating interference as noise (TIN). From Fig.~\ref{fig:SINR}, we observe that the performance of Benchmark~1 under TIN quickly saturates as the interference dominates the noise for both users. Moreover, we see from Fig.~\ref{fig:SINR} that the proposed method outperforms Benchmark~2 for both TIN and NOMA suggesting that, for the considered setting with large RIS and large wall apertures, separating users in the spatial domain is more advantageous. However, we note that, as the transmit power increases eventually interference dominates, and the sum rate of the proposed method with TIN saturates too. \textcolor{blue}{In comparison with} the benchmark schemes, the proposed scheme using NOMA exploits non-LOS links to generate a full-rank end-to-end channel, exploiting both spatial and power domains. In summary, Fig. \ref{fig:SINR} underlines the importance of exploiting non-LOS components in multi-user scenarios.

\section{Conclusion}
In this paper, we have introduced a new NF MIMO channel model accounting for reflections from NSs in addition to the LOS link and point scattering. Although recent measurement campaigns have shown the relevance of imperfect reflection from large NSs in the environment, the existing channel models do not account for this effect, constituting a research gap that is addressed in this paper. We have verified the accuracy of the proposed model by comparing it with the numerical evaluation of the Huygens-
Fresnel integral in \eqref{eq: Huygen}. Furthermore, via simulations, we have demonstrated that the impact of NSs is not only non-negligible but also beneficial for achieving a multiplexing gain in multi-user communication scenarios.

\appendix
%\section{Proof of Lemma~\ref{lem: tworxtx}}
\label{app: tworxtx}
Consider (\ref{eq:correlation1}) and decouple it into the following two cases:\\ %$\bu'\neq\bu$ and $\bu'=\bu$:\\
\textbf{Case 1 ($\bu'\neq\bu$):} %$[\bR]_{(n,m,n',m')}=\frac{1}{|\Uset|}$
%\begin{align}
%\label{eq:correlation2}
%    &\times\iint\limits_{\bu\in\Uset}\iint\limits_{\bu'\in\Uset}\Ex\Big\{ I(\bu,\bu_{\tx,n},\bu_{\rx,m})\Big\}\Ex\Big\{ I^*(\bu',\bu_{\tx,n'},\bu_{\rx,m'})\Big\}\dd x'\dd y'\dd x\dd y\nonumber\\
%    &=0,
%\end{align}
\begin{align}
\label{eq:correlation2}
[\bR]_{(n,m,n',m')}\!=\!\frac{1}{|\Uset|}\!\iint\limits_{\bu\in\Uset}\iint\limits_{\bu'\in\Uset}\!\!&\Ex\Big\{ I(\bu,\bu_{\tx,n},\bu_{\rx,m})\Big\}\\
&\!\!\times\Ex\Big\{ I^*\!(\bu',\bu_{\tx,n'},\bu_{\rx,m'})\Big\}\dd x'\!\dd y'\dd x\dd y,\nonumber
\end{align}
resulting in $[\bR]_{(n,m,n',m')}=0$ since each expectation inside the double integrals is zero in Regime 3.\\
\textbf{Case 2 ($\bu'=\bu$):}\\ First, let us expand the term of $\|\bu-\bu_{\rx,m'}\|$ in terms of $\|\bu-\bu_{\rx,m}\|$. Thereby, $\|\bu-\bu_{\rx,m'}\|$ is equal to the following quantity:
\begin{equation}\label{eq:29}
    \|\bu-\bu_{\rx,m}\|+\|\bu_{\rx,m}-\bu_{\rx,m'}\|\sin(\theta_\rx^l)+\bigo(\|\bu_{\rx,m}-\bu_{\rx,m'}\|^2).
\end{equation}
Note that, based on Assumption A1, we can neglect the last term in~\eqref{eq:29}. A similar expansion can be made for the Tx side by substituting the indices $\rx$ and $m$ with $\tx$ and $n$, respectively. To this end, the expression for $[\bR]_{(n,m,n',m')}$ simplifies as follows:
%final equation is simplified to $[\bR]_{(n,m,n',m')}=$
\begin{align}
 &[\bR]_{(n,m,n',m')}=\frac{1}{|\Uset|}\nonumber\\&\times\iint\limits_{\bu\in\Uset} \Ex\Big\{\e^{\jj \kappa(\|\bu_{\rx,m}-\bu_{\rx,m'}\|\sin(\theta_\rx^l)+\|\bu_{\tx,n}-\bu_{\tx,n'}\|\sin(\theta_\tx^l))}\Big\}\dd x\dd y.
\end{align}
Finally, we can omit the expectation based on Assumption A2.

%auto-ignore

%\bibliographystyle{IEEEtran}
%\bibliography{References}
\end{document}